\documentclass[twocolumn,showpacs,preprintnumbers,amsmath,amssymb,aps,prc]{revtex4-2}

\usepackage{empheq}
\usepackage{xcolor}
\usepackage{verbatim}
\usepackage{graphicx}
\usepackage{dcolumn}
\usepackage{bm}
\usepackage{lineno}
\usepackage{float}

\usepackage[colorlinks=true, allcolors=blue]{hyperref}

\usepackage[caption=false]{subfig}

\begin{document}
\title{How not to measure a false QCD critical point}
\author{Zachary Sweger}
\affiliation{Department of Physics and Astronomy, University of California, Davis, California 95616, USA}
\author{Daniel Cebra}
\affiliation{Department of Physics and Astronomy, University of California, Davis, California 95616, USA}
\author{Xin Dong}
\affiliation{Lawrence Berkeley National Laboratory, Berkeley, California 94720, USA}

\date{\today}

\begin{abstract}

Fluctuations of conserved charges are a golden channel for measuring a QCD critical point in relativistic heavy-ion collisions. These fluctuations are quantified by measuring high-order cumulants of baryon-number distributions at a given centrality. Using simulated proton-number cumulants as an example, we discuss how the correlation between particle identification and centrality measurements can distort particle-number distributions. These distortions can easily create large fluctuations in high-order cumulants that might be mistaken for a critical-point signature. We show that certain measurement choices can make the analysis more or less vulnerable to these false signals. We motivate this by considering how the two-dimensional probability space of proton-number versus multiplicity is shaped by analysis choices. We then demonstrate this vulnerability with simulated Au+Au collisions at $\sqrt{s_{NN}}=3.9$~GeV in UrQMD, and two toy models of detector responses to certain classes of events. We explain how an analyzer might observe a false critical signature, and how to avoid doing so, even in a challenging experimental environment.

\end{abstract}

\maketitle

\section{Introduction}

\begin{figure}
  \begin{center}
    \includegraphics[width=0.5\textwidth]{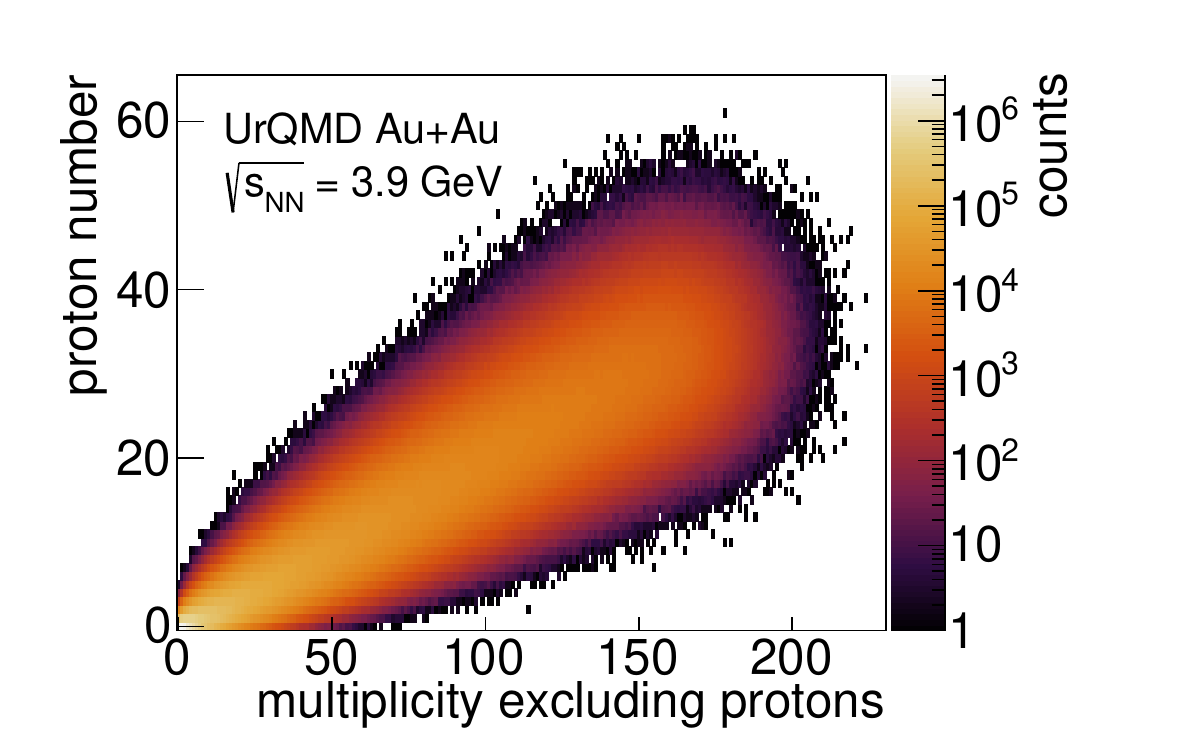}
    \caption{Distribution of proton number versus multiplicity (excluding protons) in UrQMD Au+Au collisions at $\sqrt{s_{NN}}=3.9$~GeV.}
    \label{fig:protonsvsmult}
  \end{center}
\end{figure}

In this article we highlight some of the experimental challenges associated with measuring critical fluctuations in heavy-ion collisions. In many other physics measurements, experimental backgrounds are well-understood and accounted for in the analysis: new-particle searches subtract combinatorial backgrounds; dark matter searches suppress backgrounds from minuscule contamination of radioactive nuclei. What then is an analogous experimental background for a measurement of the kurtosis of a proton-number distribution in relativistic Au+Au collisions? We demonstrate that the dominant background arises from the different ways that measurements of particle identification and centrality respond to anomalous events. By being cognizant of these backgrounds and by making strategic analysis decisions, analyzers can avoid measuring a false signal.

In relativistic heavy-ion collisions, high-order cumulants of conserved charges are expected to be sensitive to the onset of deconfinement and a critical point in the QCD phase diagram. In the vicinity of a critical point, critical fluctuations in baryon number, transverse momentum, charge, and strangeness may be quantifiable using the cumulants of distributions of these quantities. In particular, high-order cumulants of conserved charges and their ratios, such as $C_4/C_2=\kappa\sigma^2$, are predicted to have non-monotonic collision-energy dependence near a QCD critical point~\cite{PhysRevLett.102.032301, Stephanov_2011}. Under certain assumptions about the chemical-freezeout curve, a critical point may be accompanied by a large positive peak in kurtosis and $C_4/C_2$~\cite{PhysRevLett.107.052301}.

The $i$-th order cumulants, $C_i$, of baryon-number distributions and cumulant ratios have been measured in Au+Au collisions by the STAR Experiment at the Relativistic Heavy-Ion Collider~\cite{3GeVLongPaper,3GeVShortPaper,ProtonCumulantRatioOrdering,STAR27and54and200,StarBESICumulantsShortPaper,StarBESICumulants,STARDeuteronCumulants}, the HADES Collaboration at SIS18~\cite{HadesProtonCumulants}, and in Pb+Pb collisions by the ALICE Collaboration at the LHC~\cite{AliceProtonCumulants,AliceAntideuteronCumulants}. Theoretical predictions of the location of a QCD critical point are converging on the high baryon chemical potential region around $\mu_B$=500-650~MeV~\cite{PhysRevD.101.054032, GAO2021136584, PhysRevD.104.054022, refId0, sorensen2024locatingcriticalpointhadron, hippert2023bayesianlocationqcdcritical}. These chemical potentials are covered by STAR's recent fixed-target data~\cite{2019TheSB}. To date, the only proton fluctuations measurements published from the fixed-target program are at $\sqrt{s_{NN}}$=3~GeV ($\mu_B\approx720$~MeV)~\cite{3GeVLongPaper,3GeVShortPaper} and publication of the remaining datasets is pending. Understanding experimental backgrounds in this fixed-target data is crucial for interpreting any signal that may be observed.

It is common to use different detectors to measure multiplicity and identify particles of interest. In STAR analyses, event-by-event multiplicity was determined from the number of tracks in a time-projection chamber (TPC), excluding protons. Once the event has been classified by centrality, particles used in the analysis were determined by a combination of TPC and time-of-flight (TOF) detections. HADES measured net-charge in a hodoscope to determine centrality, and coincident TOF and energy-loss measurements to identify protons. ALICE used scintillators to measure centrality and energy-loss in a TPC to identify particles. In all of these measurements, a different detector was used for measuring multiplicity and for identifying particles of interest; a process we refer to as a mixed-detector approach.

\begin{figure}
  \begin{center}
    \includegraphics[width=0.4\textwidth]{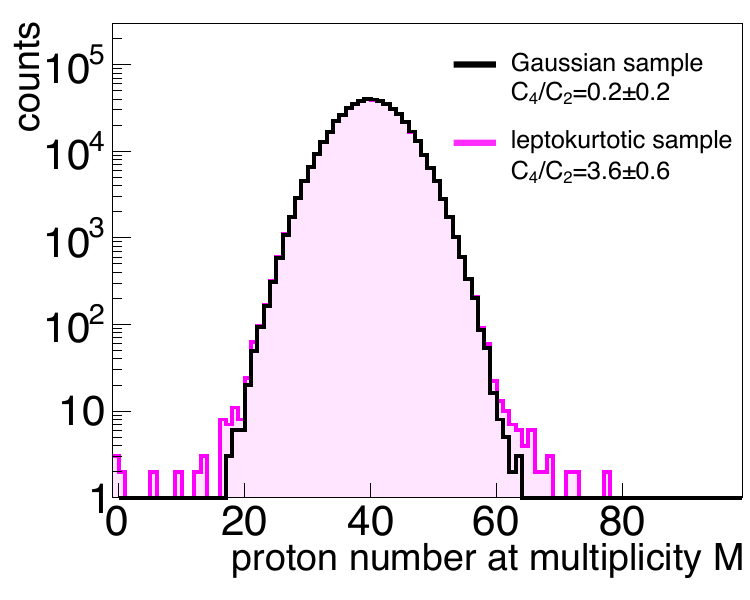}
    \caption{Two example distributions and respective $C_4/C_2$ values. The distribution outlined in black follows a normal (Gaussian) distribution. The distribution outlined in magenta is sampled from the same normal distribution, but also samples from a much broader normal distribution in 0.1\% of events.}
    \label{fig:kurtoticplot}
  \end{center}
\end{figure}

High-order cumulants are measured as a function of multiplicity and are interpreted in terms of centrality. For simplicity, we focus on the measurement of the cumulants of proton number, $n$. Rather than measuring cumulants of the probability distribution of raw proton number, $P(n)$, we instead measure cumulants of the probability distribution of proton number, given a multiplicity $M$: $P(n|M)$. The moments of one-dimensional slices in the correlation between proton number and multiplicity shown in Fig.~\ref{fig:protonsvsmult} is thus the entire measurement.  

Rare events, in which the correlation between proton number and multiplicity changes, can create large distortions in the proton number distribution at each measured multiplicity. These distortions can cause enhanced high-order cumulants. Figure~\ref{fig:kurtoticplot} shows two simulated samples of proton-number distributions for a given multiplicity $M$. In both samples the proton number is generated from normalized Gaussians with mean $\mu$, and standard deviation $\sigma$: $G(n,\mu,\sigma)$. The black distribution is sampled from $P(n|M)=G(n,\mu=40,\sigma=5)$. The magenta distribution is sampled from:
\begin{equation}
    P(n|M)=(1-\alpha)G(n,\mu,\sigma_1)+\alpha G(n,\mu,\sigma_2),
\end{equation}
with $\alpha = 0.001$, $\mu=40$, $\sigma_1=5$, and $\sigma_2=15$. In 0.1\% of events, the proton number was sampled from a Gaussian distribution with a much broader width. In other words, in 0.1\% of events, the correlation between multiplicity and proton number was much weaker. This rare and spontaneous decorrelation is enough to raise $C_4/C_2$ from being consistent with 0, to $3.6\pm0.6$. This enhancement is at a level and significance that would constitute a signal in the search for the QCD critical point. 

The key to any high-order cumulants measurement is understanding low-statistics outliers from the bulk behavior of the data. We demonstrate that the mixed-detector approach can result in rare and spontaneous changes in the correlation between multiplicity and proton number, resulting in anomalously-large high-order cumulants. Anomalous events alone are not responsible for enhanced fluctuations. Instead, the expression of anomalous events in high-order cumulants depends on analysis choices, which can be engineered to suppress these contributions. 

There are other effects that distort high-order cumulants of particle-number distributions in relativistic collisions. Imperfect mapping between event multiplicity and impact parameter distorts these cumulants. Such distortions are referred to as volume fluctuations. The effect of volume fluctuations and how to correct for them are described in Ref.~\cite{BRAUNMUNZINGER2017114,PhysRevC.88.034911}. Distortions due to detector inefficiencies are treated in Ref.~\cite{PhysRevC.95.064912}. Both of these are well-understood effects and are not explained further here.

In Sec.~\ref{section:math} we show how a different correlation between proton number and multiplicity in detector-induced fluctuations leads to enhanced high-order central moments. In Sec.~\ref{section:detectoreffects} we explore how detector responses to various classes of events change the correlation between proton number and multiplicity. In Sec.~\ref{section:numericalanalysis} we introduce two toy models of detector responses to UrQMD Au+Au collisions at $\sqrt{s_{NN}}=3.9$~GeV in order to demonstrate this change in correlation. The first toy model simulates out-of-time pileup in order to investigate how using fast and slow detectors in the presence of pileup can enhance or suppress detector-induced fluctuations. The second toy model simulates the effect of a detector undergoing fluctuations in acceptance. The results of these toy models are evaluated in Sec.~\ref{section:results}. In Sec.~\ref{section:conclusions}, we summarize these findings in a short list of prescriptions for fluctuations analyses. These prescriptions describe how to avoid measuring enhanced high-order cumulants and thus a false QCD critical-point signature.

\section{Anomalous Fluctuations and High-Order Moments}
\label{section:math}

\begin{figure}
  \begin{center}
    \includegraphics[width=0.4\textwidth]{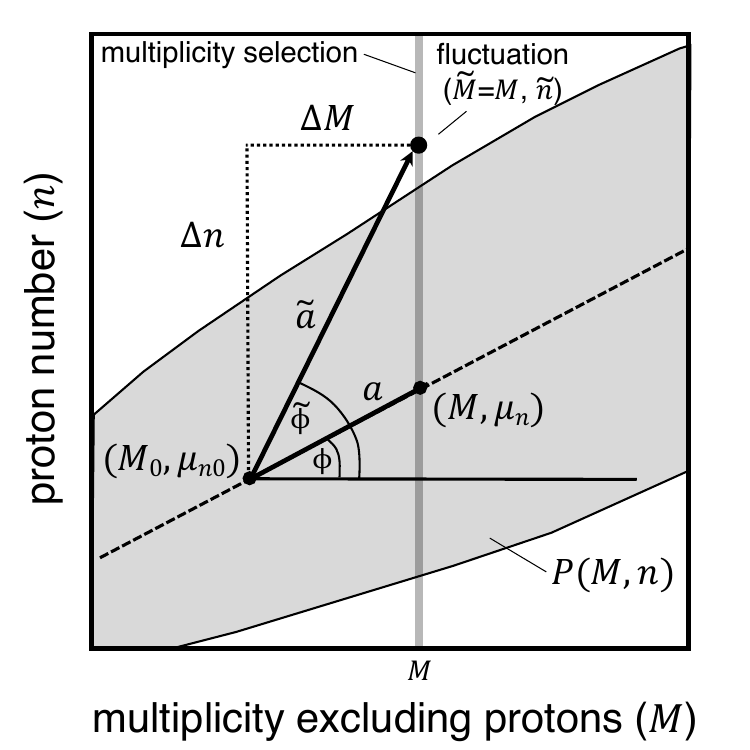}
    \caption{A zoomed-in view of a fluctuation embedded in the proton-number and multiplicity probability distribution used to clarify the variables in Sec.~\ref{section:math}. The proton-number cumulants are calculated at multiplicity $M$, and a test fluctuation is shown at $(\tilde{M}=M,\tilde{n})$.}
    \label{fig:theoryplot3}
  \end{center}
\end{figure}

In this section we evaluate the impact of detector-induced fluctuations on central moments. We demonstrate that the effect of a fluctuation on high-order central moments is determined by the correlation between proton number and multiplicity in the anomalous event. 

Cumulants of a distribution are expressed in terms of the central moments up to sixth order as:
\begin{align}
\begin{split}
    C_1 &=\mu_1 \\
    C_2 &=\mu_2 \\
    C_3 &=\mu_3 \\
    C_4 &=\mu_4 - 3\mu_2^2 \\
    C_5 &=\mu_5 - 10\mu_2\mu_3 \\
    C_6 &=\mu_6 + 30\mu_2^3 - 15\mu_2\mu_4 - 10\mu_3^2.
\end{split}
\end{align}

The $i$\textsuperscript{th}-order central moment of the proton number $n$ at some multiplicity $M$ is given by: 
\begin{equation}
\mu_i(M) =\langle(\delta n | M)^i\rangle = \int(n-\mu_{n})^iP(n|M)dn,
\end{equation}
where $\mu_{n}$ is the mean proton number given $M$, $\delta n = n-\mu_{n}$, and $P(n|M) = {P(M,n)}/{P(M)}$. We can evaluate the impact of rare detector-induced fluctuations from the truth probability distribution by separating out the contributions to $\mu_i$ from such fluctuations. We first expand the measured probability distribution $P(M,n)$:
\begin{equation}
P(M,n) = (1-\alpha)P(M,n)_{\text{truth}} +\alpha P(M,n)_{\text{fluc.}},
\end{equation}
where $\alpha$ is the fraction of events with detector-induced fluctuations.

Separating the probability in this way allows us to separate contributions to the moments:
\begin{equation}
\mu_{i} = (1-\alpha)\bar{\mu}_{i} +\alpha\tilde{\mu}_{i},
\end{equation}
where $\bar{\mu}_{i}$ is the $i$\textsuperscript{th}-order moment of the truth probability distribution, and $\tilde{\mu}_{i}$ is the contribution to the measured moment $\mu_{i}$ from detector-induced fluctuations. 

We aim to quantify how a detector-induced fluctuation that changes the measured proton number and multiplicity can distort high-order moments of the multiplicity bin in which it is measured. For simplicity, we consider an event with multiplicity $M_0$ and proton number $n=\mu_{n0}$, where $\mu_{n0}$ is the mean proton number at $M_0$. We take a fluctuation that changes its proton number to $\tilde{n}=\mu_{n0}+\Delta n$ and its multiplicity to $\tilde{M}=M_0+\Delta M$. These selections result in a fluctuation in a single bin located at $(\tilde{M},\tilde{n})$, as pictured in Fig.~\ref{fig:theoryplot3} and represented by the probability distribution:
\begin{equation}
P(M,n)_{\text{fluc.}} = \delta(M-\tilde{M},n-\tilde{n}).
\end{equation}
The deviation from ($M_0$,$\mu_{n0}$) has magnitude $\tilde{a} = \sqrt{(\Delta n)^2 + (\Delta M)^2}$ and angle $\tilde{\phi}$, such that $\tilde{n}=\mu_{n0}+\tilde{a}\sin\tilde{\phi}$. We refer to $\tilde{\phi}$ as the fluctuation angle. In these coordinates, the fluctuation is expressed as: 
\begin{equation}
P(M,n)_{\text{fluc.}} = \delta(M-\tilde{M},n-\mu_{n0}-\tilde{a}\sin\tilde{\phi}).
\end{equation}

The contribution of this fluctuation to the $i$\textsuperscript{th} moment in the multiplicity bin $M=\tilde{M}$ is then:

\begin{equation}
\label{eq:muicontribution}
\tilde{\mu}_{i}(M) = (\mu_{n0}+\tilde{a}\sin\tilde{\phi}-\mu_n)^i.
\end{equation}

The mean of the proton-number distribution at multiplicity $M$ occurs at $(M,\mu_{n})$. Relative to $(M_0,\mu_{n0})$, $\mu_{n}$ can be expressed as $\mu_n=\mu_{n0}+a\sin\phi$, where $a$ is the distance between $(M_{0},\mu_{n0})$ and $(M,\mu_{n})$, and $\phi$ is the angle between them. This angle between the mean proton number at $M_0$ and the mean proton number at $M$ is referred to as the mean angle. Now Eq.~\ref{eq:muicontribution} becomes:

\begin{equation}
\tilde{\mu}_{i}(M) = (\tilde{a}\sin\tilde{\phi}-a\sin\phi)^i.
\end{equation}

\begin{figure}[h]
  \begin{center}
    \includegraphics[width=0.3\textwidth]{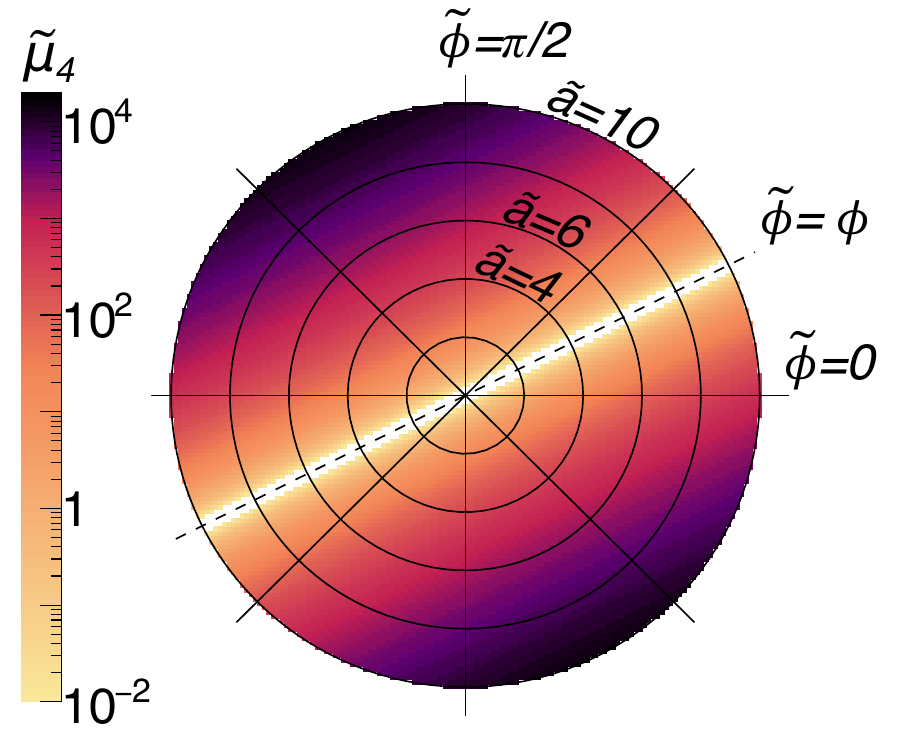}
    \caption{$\tilde{\mu}_4$ for a fluctuation with magnitude $\tilde{a}$ and angle $\tilde{\phi}$ embedded in a distribution of proton number versus multiplicity with mean angle $\phi$.}
    \label{fig:fluctuationangle}
  \end{center}
\end{figure}

Expressing $a$ in terms of the other quantities, we arrive at:
\begin{equation}
\label{eq:slamdunk}
\tilde{\mu}_{i}(M) = (\tilde{a}\sin\tilde{\phi}-\tilde{a}\cos\tilde{\phi}\tan\phi)^i.
\end{equation}

When $\tilde{\phi}=0$ the fluctuation is entirely in the multiplicity direction and $\tilde{\mu}_{i}(M) = (-\tilde{a}\tan\phi)^i$. When $\tilde{\phi}=\pi/2$ then $a=0$, and the fluctuation is entirely in the proton-number direction such that $\tilde{\mu}_{i}(M) \sim \tilde{a}^i$. When the fluctuation angle is equal to the mean angle, $\tilde{\phi}=\phi$, then $\tilde{a}=a$, and the contribution of the fluctuation to each moment vanishes. Thus, contributions from detector-induced fluctuations are suppressed when the correlation between proton number and multiplicity in the fluctuation matches the global correlation. 

To visualize the effect on $\mu_4$ of a fluctuation with magnitude $\tilde{a}$ and angle $\tilde{\phi}$, we plot, in Fig.~\ref{fig:fluctuationangle}, $\tilde{\mu}_4$ versus these two variables for some mean angle $\phi$. Enhancements to high-order moments are suppressed when the marginal enhancement (or suppression) of proton number and multiplicity in an anomalous event maintains the same ratio of proton number to multiplicity as events without detector-induced fluctuations. A detector-induced fluctuation can have a large effect on high-order moments if the ratio of proton number to multiplicity deviates from that of good events.

\section{Detector Effects on Multiplicity Correlations}
\label{section:detectoreffects}

\begin{figure}[h]
  \begin{center}
    \includegraphics[width=0.4\textwidth]{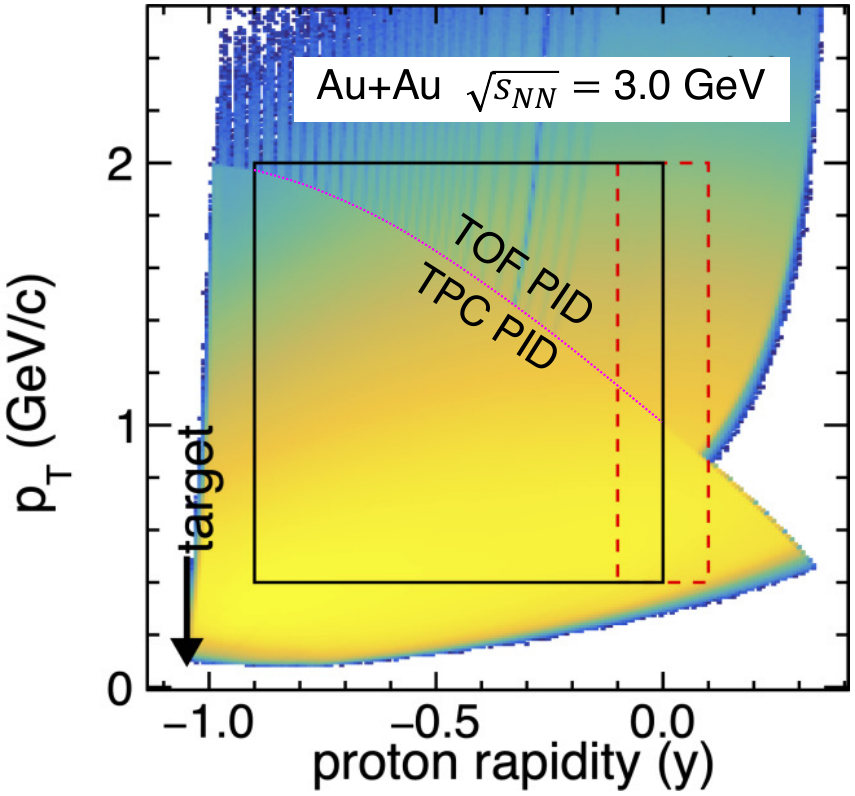}
    \caption{Analysis window used at $\sqrt{s_{NN}}=3$~GeV from the STAR Fixed-Target Program.~\cite{3GeVLongPaper, 3GeVShortPaper}}
    \label{fig:3gevacceptance}
  \end{center}
\end{figure}

The detectors used for multiplicity and PID measurements define the fluctuation angle $\tilde{\phi}$. The nontrivial impact of detector efficiency on cumulants was treated in Ref.~\cite{PhysRevC.95.064912}. These efficiency considerations as well as the natural multiplicity and proton number define the mean angle $\phi$. In addition to efficiency considerations, different detectors respond to certain classes of events in different ways. These discrepant responses to various event classes have the potential to cause cumulants of event-wise distributions to have large values that can be mistaken as a signal. To illustrate this point, we introduce two toy models of the measurement of proton-number cumulants:
\begin{enumerate}
    \item fast and slow detectors in the presence of out-of-time pileup
    \item two detectors, one stable and one with unstable acceptance
\end{enumerate}

These examples are illustrative of the problems that can arise from using different detectors for multiplicity and PID. Both examples use proton-number cumulants as a case study, but the conclusions are broadly applicable to other cumulants measurements. The examples illustrate that enhanced fluctuations occur not simply due to the existence of pileup, acceptance fluctuations, or other anomalous events. These fluctuations are caused specifically by anomalous events detected with a mixed-detector approach. The differing response of detectors to anomalous events changes the fluctuation angle and enhances high-order central moments and cumulants.

\subsection{Example 1: Pileup with fast and slow detectors}
\label{section:tofandpileup}
In Beam-Energy-Scan I, the STAR Experiment used a time-projection chamber for measuring the event-by-event multiplicity. TPCs are relatively slow detectors, as the measurable event rate is limited by the drift time of charged tracks. The STAR TPC has a drift time of 40~$\mu$s~\cite{STAR:2002eio}. Time-of-flight detectors, on the other hand, use picosecond-scale timing resolution to identify particles.

With the slow drift time of TPCs, it is common for a second collision to occur while ionization from the first collision is still being collected. If tracks from this second collision are not separated from tracks from the first collision, the second collision is referred to as pileup. In accelerators such as RHIC and the LHC, radio-frequency (RF) oscillations in the electromagnetic field define discreet buckets in which ions can maintain closed orbits around the ring. Each of these buckets can be independently filled with ions or left empty. A filled bucket is referred to as a bunch. 

Pileup can be categorized broadly as either in-time or out-of-time. In-time pileup occurs when there are multiple collisions within the same bunch crossing. The time elapsed between the two collisions is small, so fast detectors like TOF and silicon trackers may measure particles from each collision. However hits from in-time pileup in a TOF detector will often have time-of-flight values that are skewed such that these hits are not identified with any particle.  Out-of-time pileup happens when a collision from one bunch crossing triggers the detectors, followed by a collision in a subsequent bunch crossing, while information from the first collision is still being collected. In a slow detector like a TPC, out-of-time pileup tracks may be counted as part of the triggered event. Fast detectors typically have short collection-time windows following a trigger, and do not detect out-of-time tracks.

In STAR's recent publications of proton cumulants at $\sqrt{s_{NN}}=3$~GeV in the Fixed-Target Program, the total pileup rate was 0.46\%~\cite{3GeVLongPaper, 3GeVShortPaper}. Rather than reject pileup, an unfolding method was applied in order to correct for the effect of pileup on the cumulants~\cite{NONAKA2020164632,ZHANG2022166246}.

The unfolding method assumes that the detector has the same efficiency for pileup tracks, so that a pileup event is a simple sum of two individual events. This assumption starts to break down in the mixed-detector approach. When a fast detector like TOF is used for PID, much of the in-time and out-of-time pileup is sufficiently out-of-time that those pileup protons are not identified. However a slow detector like a TPC will include pileup tracks in the multiplicity. In the mixed-detector approach, many pileup events are seen as a single collision by a fast detector, and as a double collision by a slow detector.

In STAR's previous analyses of proton-number cumulants, protons are measured using both the TPC and TOF. Energy-loss in the TPC is used to identify protons up to a given threshold momentum, above which TOF is used in order to maintain proton purity. The analysis window used at $\sqrt{s_{NN}}=3$~GeV is shown in Fig.~\ref{fig:3gevacceptance}. For particles with momenta greater than 2~GeV, a hit in the TOF with a mass cut was required. When there is out-of-time pileup, the TPC measures the multiplicity as the sum of charged tracks from both collisions. Since both the TPC and TOF are used to identify protons, the proton number is made up of tracks above the momentum threshold from a single collision, and tracks below the momentum threshold from both collisions. These out-of-time pileup events will thus have an abnormally small number of protons, given their centrality as measured by the TPC. 

If proton number and multiplicity measurements are both performed by a slow detector, then pileup tracks are counted in both. In this case, contributions to the cumulants from pileup are suppressed due to the correlation between $n$ and $M$. When pileup is observed in both numbers, the fluctuation angle matches the mean angle. Moreover, when $n$ and $M$ both contain pileup, the remaining contributions to the cumulants from pileup can be corrected by the unfolding method introduced in Ref.~\cite{NONAKA2020164632,ZHANG2022166246}.

If proton number is measured using a fast detector and multiplicity is measured by a slow detector, then out-of-time pileup will contribute to a long low-proton-number tail, which skews the high-order cumulants. Finally, if proton number is measured using a slow detector and multiplicity is measured by a fast detector, then out-of-time pileup will contribute to a long high-proton-number tail, which also skews high-order cumulants. 

\subsection{Example 2: Unstable acceptance}

\begin{figure}
  \begin{center}
    \includegraphics[width=0.45\textwidth]{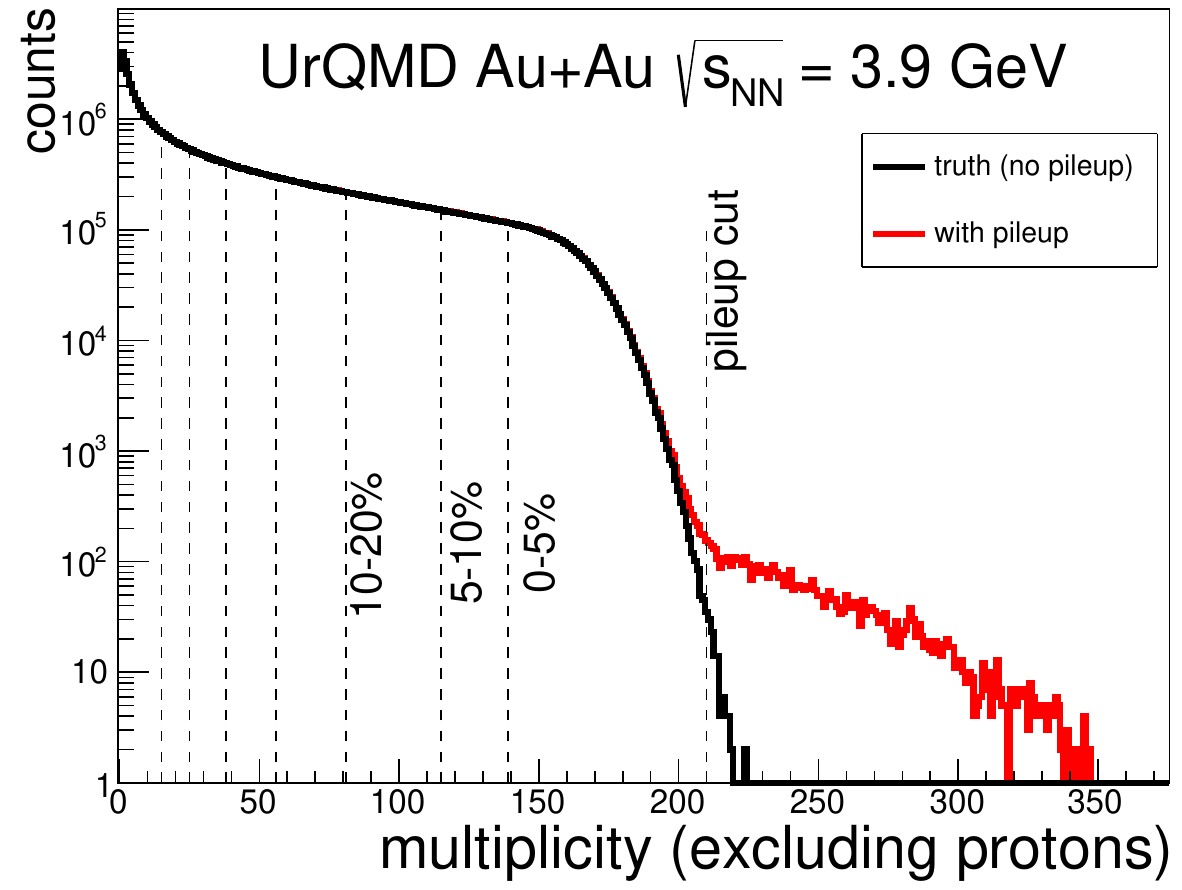}
    \caption{Multiplicity distributions excluding protons, with and without pileup tracks. Centrality cuts are shown as dashed lines up to 60\% centrality, including a pileup cut at the upper edge of the 0-5\% bin.}
    \label{fig:refmult}
  \end{center}
\end{figure}

Another way to see the impact of the mixed-detector approach is to examine events in which one detector has an unstable acceptance and the other does not. For example, this can happen if one of the detectors is segmented into several active areas that independently reboot during a run. Acceptance can also fluctuate if the algorithms used by a detector are prone to failure. Many time-of-flight detectors rely on a start-time algorithm to calculate exactly when each collision occurred. Should that algorithm fail for some subset of events, the efficiency for correct proton identification might fall or drop to zero within the TOF acceptance.

If an unstable detector is used to measure both PID and multiplicity, then events when the detector is partially or fully inactive have reduced multiplicity, and reduced identifiable protons. In this case, the proton number scales with the reduced apparent centrality and the outlier events ``blend in" with the bulk. If an unstable detector is used to identify protons while the stable detector measures centrality, then bad events can result in very central collisions with few identified protons. Finally if the stable detector identifies protons but the unstable detector measures centrality, then central events will be sorted as more peripheral, resulting in high proton-number tails.

\section{Numerical Analysis}
\label{section:numericalanalysis}

\begin{figure}
  \begin{center}
    \includegraphics[width=0.5\textwidth]{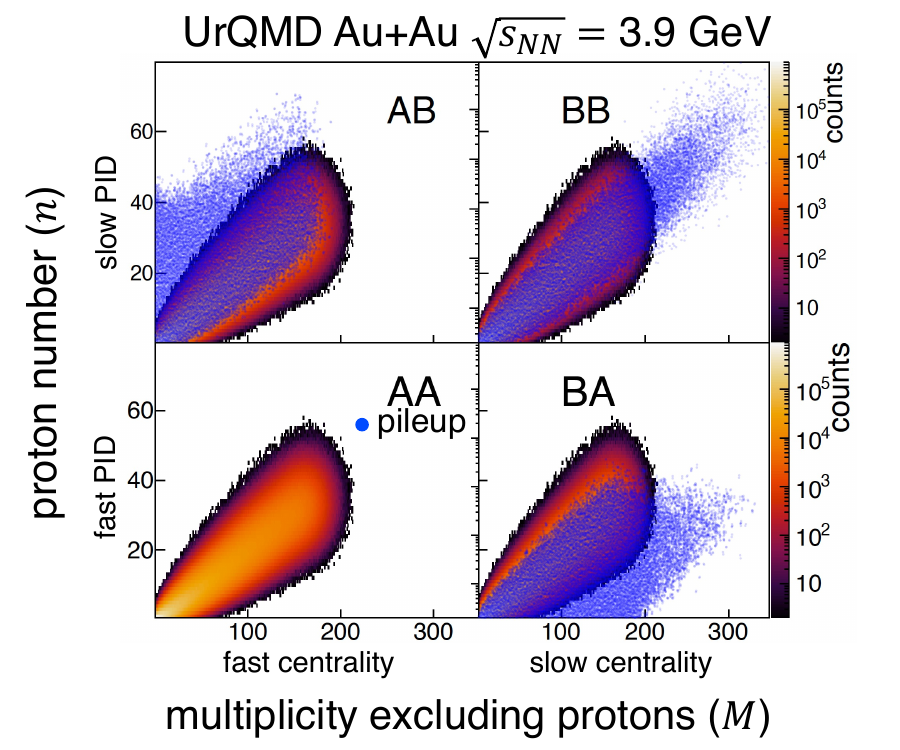}
    \caption{Distribution of detected proton number ($n$) versus multiplicity ($M$) for the four fast-slow detector combinations where the pileup events are shown in blue. The pileup rate is 0.2\%.}
    \label{fig:cokurtosismodel1}
  \end{center}
\end{figure}

In this section we introduce two toy models of detector responses to UrQMD events in order to demonstrate the risks of a mixed-detector approach. Both toy models use $6.4\times10^7$ Au+Au collisions at $\sqrt{s_{NN}}=3.9$~GeV from UrQMD in cascade mode. This energy was chosen because it corresponds to one of the datasets collected by the STAR Experiment's Fixed-Target Program. We use the same analysis window that was used in the analysis of the $\sqrt{s_{NN}}=3$~GeV data: $0.4<p_T<2$~GeV/c and $-0.5<y-y_{CM}<0$. 

Multiplicity distributions from these simulated events are integrated to define centrality bins. The multiplicity was defined to exclude protons in order to avoid autocorrelations. We simulate a realistic fixed-target acceptance by defining the multiplicity as all charged pions and kaons with $p_T>0.06$~GeV/c, which, after boosting to the lab frame, have pseudorapidities of $0<\eta<2.15$.

\begin{figure}
  \begin{center}
    \includegraphics[width=0.5\textwidth]{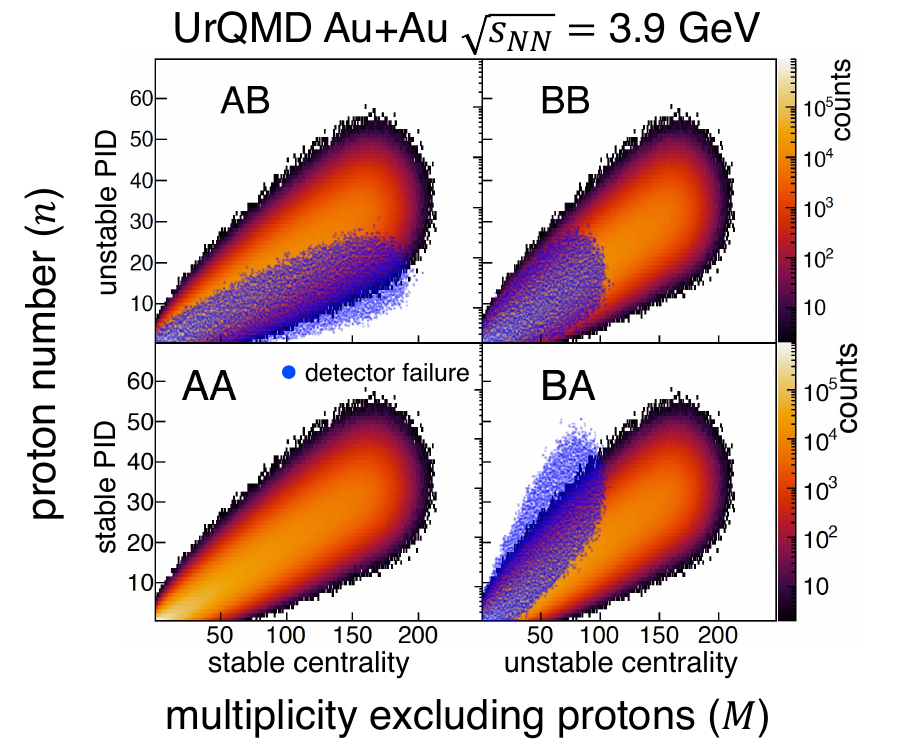}
    \caption{Distribution of detected proton number versus multiplicity for the four unstable-stable detector combinations. The 1\% of events in which the unstable detector was half-dead are shown in blue.}
    \label{fig:cokurtosismodel2}
  \end{center}
\end{figure}

The centrality bins used here are 0-5\%, 5-10\%, 10-20\%, 20-30\%, 30-40\%, 40-50\%, and 50-60\%. The multiplicity distributions with and without pileup are shown in Fig.~\ref{fig:refmult} along with the centrality cuts. Toy model~1 simulates the effects of pileup on the proton-number cumulants. In order to replicate experimental conditions, the pileup cut on multiplicity shown in Fig.~\ref{fig:refmult}, is used to reduce pileup contributions. Centrality cuts defined by integrating the red and black distributions (with and without pileup) were equivalent.

\begin{figure*}
  \centering
    \includegraphics[width=1.0\textwidth]{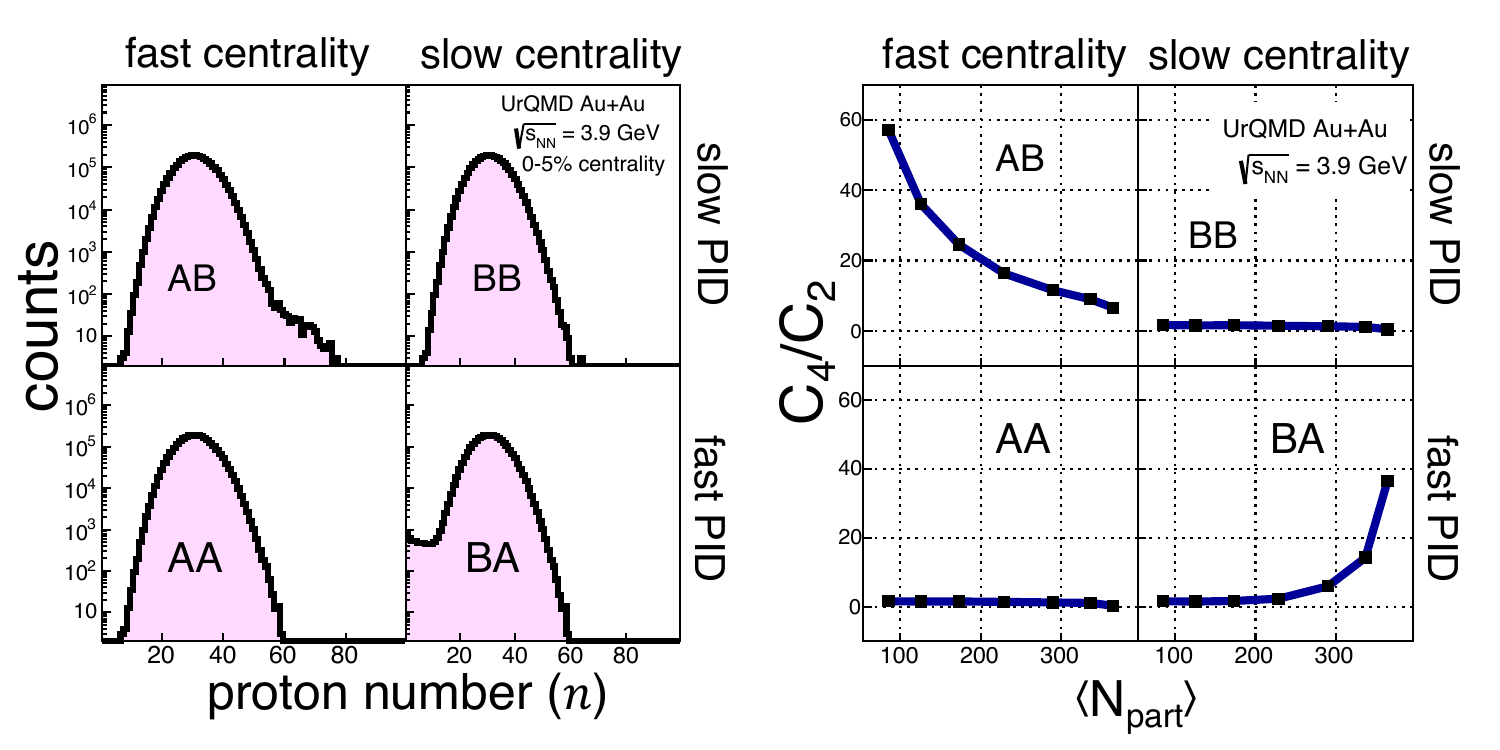}
    \caption{(left) Distribution of proton number in the 0-5\% centrality bin with a 0.2\% out-of-time pileup rate for all four combinations of fast-slow detector combinations. (right) $C_4/C_2$ as a function of centrality (plotted as $\langle N_{\text{part}}\rangle$) for each fast-slow detection method.}
    \label{fig:diagonalsmodel1}
\end{figure*}

\subsection{Toy model 1: out-of-time pileup}

In the first toy model, out-of-time pileup was simulated by sampling two collisions in $0.2\%$ of events. We simulate the responses of a fast and a slow detector to this pileup. Fast and slow toy detectors have 100\% efficiency for in-time collisions. For out-of-time pileup, the slow detector has 100\% efficiency, while the fast detector has 0\% efficiency.

The correlation between proton number and multiplicity in UrQMD events is shown in Fig.~\ref{fig:cokurtosismodel1}. In panel {\fontfamily{phv}\selectfont{\small AA}}, the fast detector is used to measure both protons and multiplicity, so no pileup tracks are included in the track sums. In panel {\fontfamily{phv}\selectfont{\small BB}}, the slow detector is used to measure everything, so that pileup events have both a high proton number and a high multiplicity. Panel {\fontfamily{phv}\selectfont{\small AB}} relies on the fast detector to measure multiplicity and the slow detector to measure protons, leading to many events with an abnormally large proton number given their multiplicity. Panel {\fontfamily{phv}\selectfont{\small BA}} shows the use of the slow detector to measure multiplicity and the fast detector to measure protons, so the pileup events often have too few identified protons for their multiplicity. The decorrelation between proton number and multiplicity in panels {\fontfamily{phv}\selectfont{\small AB}} and {\fontfamily{phv}\selectfont{\small BA}} contributes to large signals in the proton-number cumulants in these mixed-detector approaches. The correlation in panel {\fontfamily{phv}\selectfont{\small BB}} suppresses the impact of pileup on the cumulants. This result is quantified in Sec.~\ref{section:results}.

\subsection{Toy model 2: Unstable acceptance}

The second model demonstrates the effect of a detector with unstable acceptance. In this scenario two detectors are again used in tandem to measure the proton number and multiplicity, however one of the detectors is unstable. For 1\% of events, in azimuth half of the unstable detector does not detect anything, while the other half is always active. The stable detector is always 100\% active.

The correlation between proton number and multiplicity in this second toy model is shown in Fig.~\ref{fig:cokurtosismodel2}. In panel {\fontfamily{phv}\selectfont{\small AA}}, the stable detector is used to measure both protons and multiplicity, so we measure the true proton number versus multiplicity. In panel {\fontfamily{phv}\selectfont{\small BB}}, the unstable detector is used to measure both, so that in the 1\% of events in which the unstable detector fails, both proton number and multiplicity are reduced by half. The fluctuation angle matches the mean angle; thus the unstable events do not stand out from the true distribution.

\begin{figure*}
  \begin{center}
    \includegraphics[width=\textwidth]{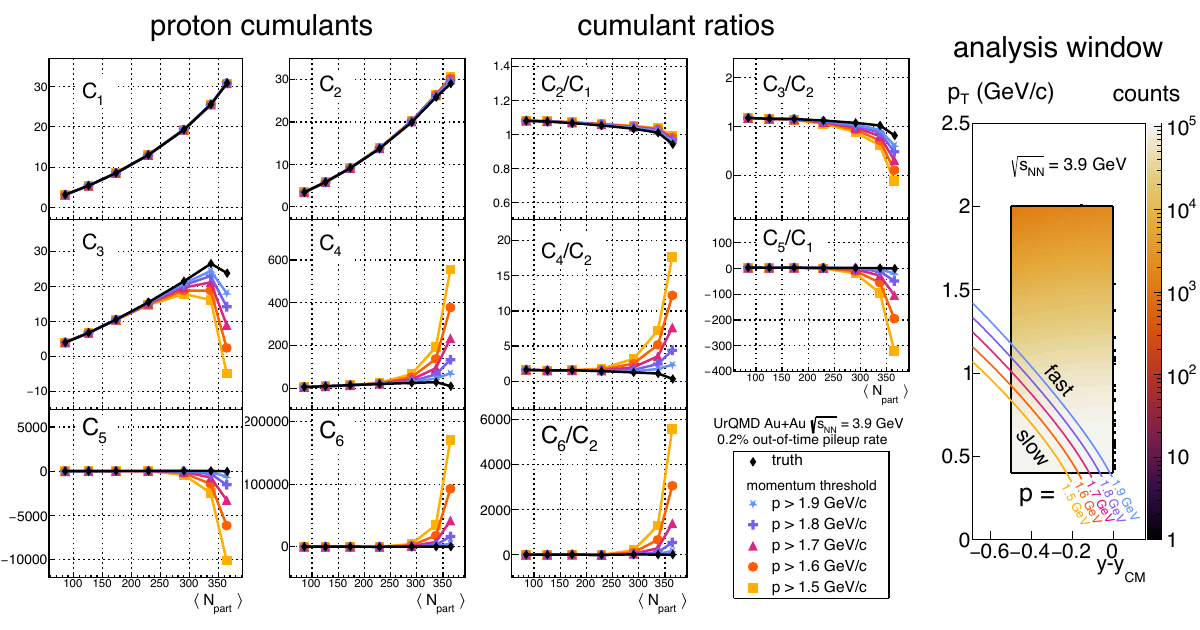}
    \caption{(left) Proton-number cumulants and cumulant ratios at $\sqrt{s_{NN}}=3.9$~GeV with varying momentum thresholds for requiring a fast detector for PID, and using a 0.2\% out-of-time pileup rate. High-order cumulants experience instability as the momentum threshold changes. (right) Analysis window with the various momentum thresholds.}
    \label{fig:cumulantsmodel1}
  \end{center}
\end{figure*}

\begin{figure}
  \begin{center}
    \includegraphics[width=0.49\textwidth]{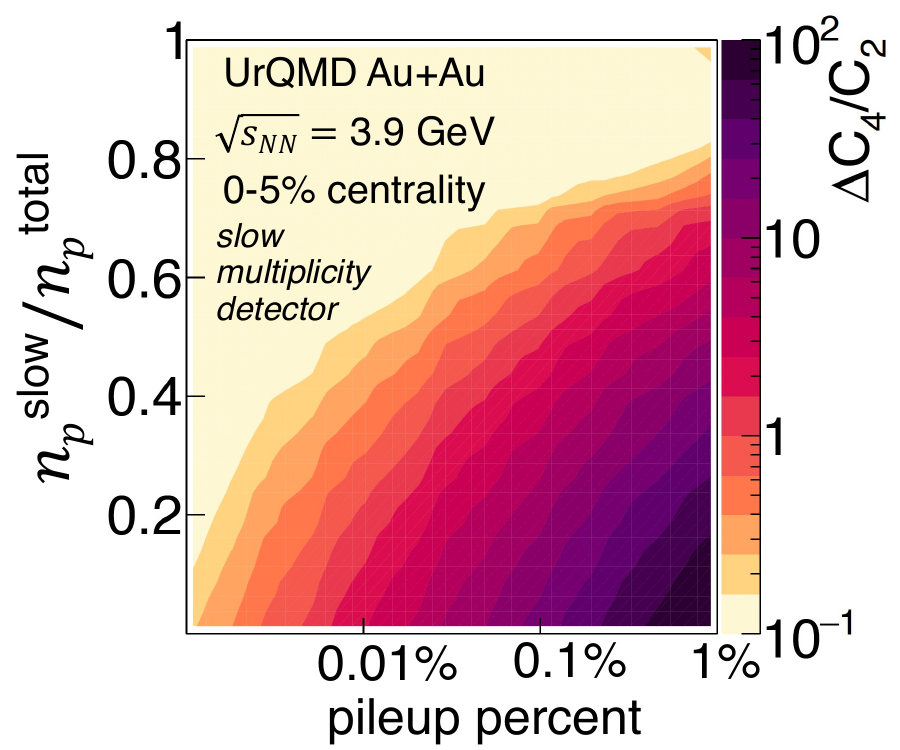}
    \caption{Enhancement of $C_4/C_2$ with several pileup rates and slow-detector acceptance fractions when using a slow detector to measure multiplicity. The axis $n_p^{\text{slow}}/n_p^{\text{total}}$ is the fraction of protons identified by the slow detector.}
    \label{fig:2dscan}
  \end{center}
\end{figure}

In panels {\fontfamily{phv}\selectfont{\small AB}} and {\fontfamily{phv}\selectfont{\small BA}}, only one of the two measurements is susceptible to the acceptance fluctuations. When taking vertical slices at a given multiplicity, this results in proton-number tails on either side of the true distributions. The impact of these tails on the cumulants is quantified in Sec.~\ref{section:results}.

\section{Results}
\label{section:results}

In this section, we evaluate the cumulants of the proton-number distributions produced in the toy models described above. 

\subsection{Toy model 1: out-of-time pileup}

The left-hand side of Fig.~\ref{fig:diagonalsmodel1} shows proton-number distributions in the toy model in the 0-5\% centrality bin, as defined by cuts in multiplicity. The right-hand side shows the resulting $C_4/C_2$ from 0\% to 60\% centrality, plotted as a function of the average number of participants $\langle N_{\text{part}}\rangle$.

Panel {\fontfamily{phv}\selectfont{\small AA}} (left) shows the proton-number distribution with no out-of-time pileup effect in the 0-5\% centrality bin. In the toy model, this is identical to the true (single collision) distribution and results in values of $C_4/C_2$ as a function of centrality near 1 (panel {\fontfamily{phv}\selectfont{\small AA}} right). Panel {\fontfamily{phv}\selectfont{\small BB}} (left) shows the proton-number distribution when both the multiplicity and the proton number are enhanced by pileup. Tails on the proton-number distribution are suppressed because the fluctuation angle is identical to the mean angle. High-order cumulants in panel {\fontfamily{phv}\selectfont{\small BB}} are distorted by pileup, but these effects are minimal and can be corrected using the unfolding approach introduced in Ref.~\cite{NONAKA2020164632,ZHANG2022166246}. These corrections assume that pileup events can be expressed as a superposition of two events, an assumption that is often invalid in the mixed-detector approach.

Panel {\fontfamily{phv}\selectfont{\small BA}} (left) shows the distribution when events with pileup-enhanced multiplicity do not have similarly enhanced proton numbers. In this case, pileup of two-midcentral collisions may be classified as a single central collision, but the fast detector identifies protons from only one of these collisions. This leads to a long low-proton-number tail.

Panels {\fontfamily{phv}\selectfont{\small BB}} and {\fontfamily{phv}\selectfont{\small BA}} are two extremes of the options for PID. It is often the case that PID is performed using several detectors. In previous STAR analyses, PID was performed with a TPC for low momenta particles, and additionally required TOF for high-momenta particles~\cite{3GeVLongPaper,3GeVShortPaper,ProtonCumulantRatioOrdering}. A more representative distribution would be intermediate between panels {\fontfamily{phv}\selectfont{\small BB}} and {\fontfamily{phv}\selectfont{\small BA}} (left). In the past, pileup was then corrected, assuming the distribution in panel {\fontfamily{phv}\selectfont{\small BB}} (left). This is problematic due to potentially large contributions to the proton-number tails caused by TOF identification. Pileup events are no longer a simple sum of two single collisions, and the standard pileup correction~\cite{NONAKA2020164632,ZHANG2022166246} should not be used.

\begin{figure*}
  \begin{center}
    \includegraphics[width=1.0\textwidth]{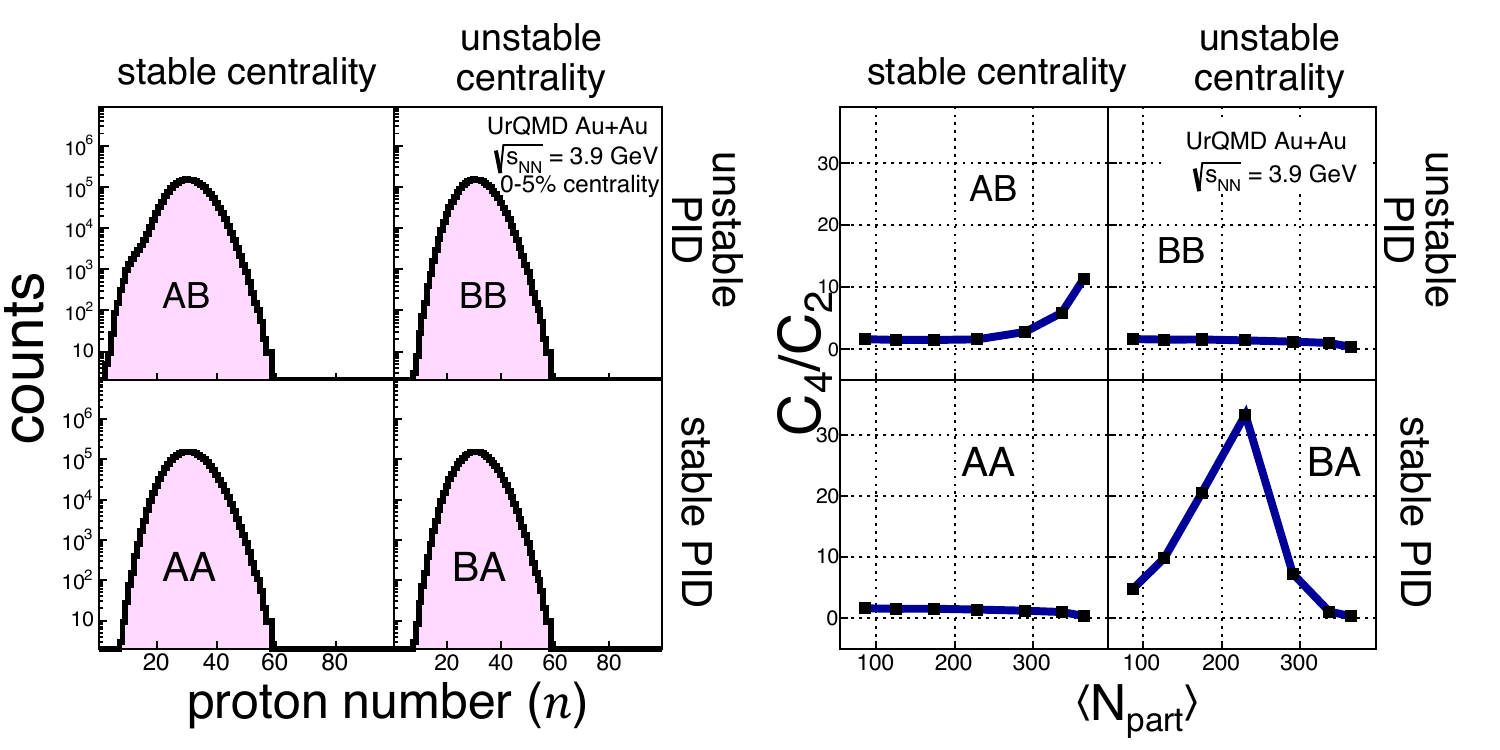}
    \caption{(left) Distribution of proton number in the 0-5\% centrality bin when the stable and unstable detectors were used to measure proton number and centrality. (right) $C_4/C_2$ in each centrality bin plotted as a function of $\langle N_{\text{part}}\rangle$ for each stable-unstable detector combination.}
    \label{fig:diagonalsmodel2}
  \end{center}
\end{figure*}

It is tempting to suggest that enhanced multiplicity from pileup could be rectified by using a fast detector to measure multiplicity, shown in panels {\fontfamily{phv}\selectfont{\small AA}} and {\fontfamily{phv}\selectfont{\small AB}} (left). However, if a slow detector is used to measure proton number, as in panel {\fontfamily{phv}\selectfont{\small AB}}, the result can be pileup events registering a low multiplicity and a high proton-number tail. It is often the case that both a fast and slow detector are used to identify protons. In this case, measuring multiplicity with a fast detector will result in something between {\fontfamily{phv}\selectfont{\small AA}} and {\fontfamily{phv}\selectfont{\small AB}} (left).

To simulate a more realistic detector environment we next examined the effect of using a slow detector to measure low-momentum protons, and a fast detector for high-momentum protons. In this model a slow detector measures the multiplicity. We choose five different momentum thresholds above which we require a fast detector for PID. The thresholds, $p=$1.5, 1.6, 1.7, 1.8, and 1.9 GeV/c, are shown on the right side of Fig.~\ref{fig:cumulantsmodel1}. The cumulants and cumulant ratios for each of these acceptance maps are displayed on the left side of Fig.~\ref{fig:cumulantsmodel1}. For $C_3$ and above, the cumulants are unstable and have strong dependence on the chosen momentum threshold. Counter to conventional wisdom, the cumulants approach their true values (shown in black) when more pileup is allowed in the proton analysis window. This is because increasing pileup in the analysis window corresponds to the fluctuation angle approaching the mean angle (as in Fig.~\ref{fig:cokurtosismodel1}, from panel {\fontfamily{phv}\selectfont{\small BA}} to {\fontfamily{phv}\selectfont{\small BB}}).

We also investigated the pileup-rate dependence of the measured kurtosis. Scanning pileup rates from 0.001\% to 1\%, we simulated various momentum thresholds above which a fast detector was used for proton identification. For each momentum threshold we calculated the fraction of protons identified by the slow detector. The enhancement of $C_4/C_2$ for each slow-acceptance fraction and each pileup rate is plotted in Fig.~\ref{fig:2dscan}. We find that up to a 1\% pileup fraction, using the slow-detector to identify $\approx$80\% of protons results in no significant enhancement of $C_4/C_2$. When the slow detector is not used for PID, the enhancement in the measured $C_4/C_2$ reaches $\approx10^2$.

\subsection{Toy model 2: Unstable acceptance}

\begin{figure*}
  \begin{center}
    \includegraphics[width=\textwidth]{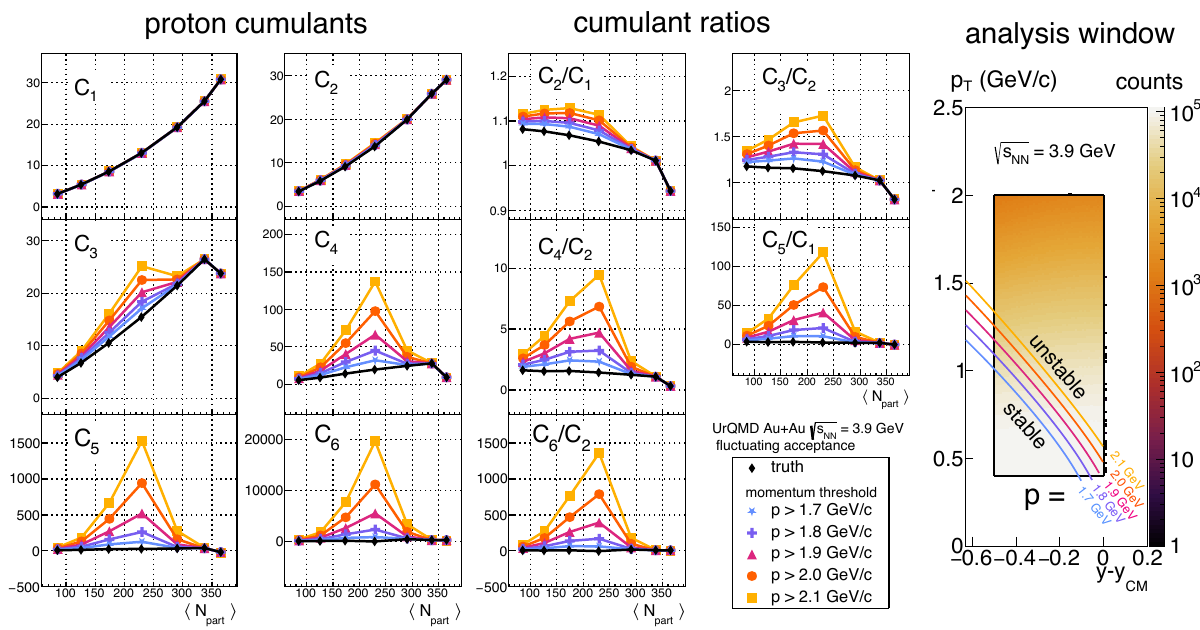}
    \caption{(left) Proton-number cumulants and cumulant ratios at $\sqrt{s_{NN}}=3.9$~GeV with varying momentum thresholds for requiring an unstable detector for PID. High-order cumulants are unstable as the momentum threshold changes. (right) Proton analysis window with momentum thresholds superimposed.}
    \label{fig:cumulantsmodel2}
  \end{center}
\end{figure*}

The left-hand side of Fig.~\ref{fig:diagonalsmodel2} shows the distribution of protons for 0-5\% centrality when the stable and unstable detectors are used to measure centrality and proton number. Panel {\fontfamily{phv}\selectfont{\small AA}} is the true distribution and {\fontfamily{phv}\selectfont{\small BB}} exhibits no obvious modification. Panel {\fontfamily{phv}\selectfont{\small BA}} is not modified because the events in which the centrality detector was unstable were shifted out of the 0-5\% centrality bin and into more peripheral bins (see Fig.~\ref{fig:cokurtosismodel2}). Panel {\fontfamily{phv}\selectfont{\small AB}} has a tail from the spontaneous failure of the proton-number measurement. Although panel {\fontfamily{phv}\selectfont{\small BA}} (left) for the 0-5\% centrality is unmodified, we can see from the right-hand side of Fig.~\ref{fig:diagonalsmodel2} that there is significant modification of $C_4/C_2$ for the 10-60\% centrality range.

On the right-hand side of Fig.~\ref{fig:diagonalsmodel2}, panel {\fontfamily{phv}\selectfont{\small BB}} does not exhibit strong deviations from {\fontfamily{phv}\selectfont{\small AA}} at any centrality. Panel {\fontfamily{phv}\selectfont{\small AB}} has a centrality dependence similar to Panel {\fontfamily{phv}\selectfont{\small BA}} of Fig.~\ref{fig:diagonalsmodel1} from model 1. This is because the low-proton-number tail is more pronounced at large centralities in both models. Panel {\fontfamily{phv}\selectfont{\small BA}} on the right of Fig.~\ref{fig:diagonalsmodel2} has very different behavior. The $C_4/C_2$ is unchanged for central collisions, but it exhibits a steep rise with falling centrality until it reaches a maximum deviation in the 20-30\% centrality bin, and then drops again. This can be understood from panel {\fontfamily{phv}\selectfont{\small BA}} in Fig.~\ref{fig:cokurtosismodel2} because all the anomalous events are shifted away from the most central multiplicities. The sudden drop in efficiency for the multiplicity detection results in the categorization of many central events as mid-central, and yet they have too many protons. This leads to a long high-proton-number tail for mid-central collisions and a correspondingly-large $C_4/C_2$.

We now examine a more realistic detector scenario in which one detector is used for proton measurements below a certain momentum threshold and another is used above the threshold. For our proton measurements, we use the unstable detector above the momentum threshold, and the stable detector below the momentum threshold. 

\begin{figure}
  \begin{center}
    \includegraphics[width=0.49\textwidth]{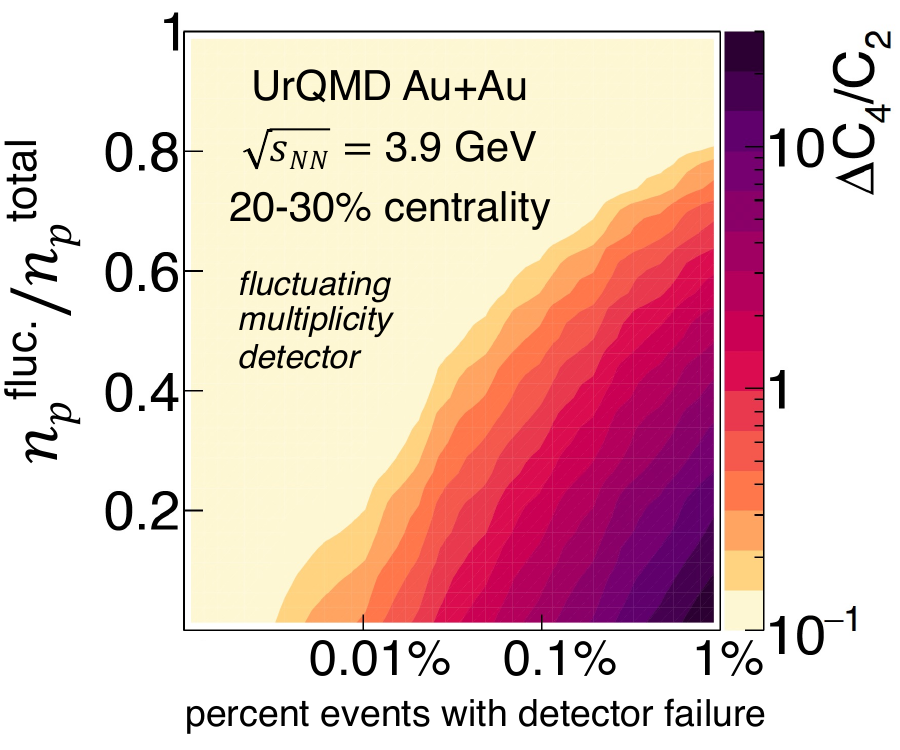}
    \caption{Enhancement of $C_4/C_2$ for 20-30\% central collisions with several detector failure rates when using a fluctuating detector for centrality. The axis $n_p^{\text{fluc.}}/n_p^{\text{total}}$ is the fraction of protons identified by the fluctuating detector.}
    \label{fig:2dscan_model2}
  \end{center}
\end{figure}

The analysis window with various momentum thresholds is shown on the right side of Fig.~\ref{fig:cumulantsmodel2}. 
We choose in this example to use the unstable detector for the multiplicity measurement.
The cumulants and cumulant ratios are on the left side of the figure. As the momentum threshold increases, the cumulants experience greater and greater enhancement. In other words, the more the unstable detector is used, the better our measurement gets. This result again underscores a counter-intuitive conclusion: acceptance fluctuations have the greatest impact on the measured cumulants when not reflected equally in the proton-number and multiplicity.

Figure~\ref{fig:2dscan_model2} plots the enhancement of $C_4/C_2$ as a function of the detector failure rate and the fraction of protons measured by the unstable detector. The enhancement in $C_4/C_2$ in Fig.~\ref{fig:cumulantsmodel2} is greatest in the 20-30\% centrality bin, so Fig.~\ref{fig:2dscan_model2} is chosen to plot the  $C_4/C_2$ at this centrality. In these simulations, the unstable detector is used to define centrality. We observe in this model that when an unstable detector measures multiplicity, it is beneficial to also maximize the amount of protons identified by the same unstable detector. The enhancement of $C_4/C_2$ in Fig.~\ref{fig:2dscan_model2} with various rates of detector failure is model dependent. This simulation describes a detector that spontaneously loses half of its acceptance in azimuth. The map in Fig.~\ref{fig:2dscan_model2} is specific to this model of detector failure and does not broadly describe fluctuations in detector acceptance.


\section{Conclusions}
\label{section:conclusions}

In previous measurements of cumulants, mixed-detector approaches were used to identify particles and multiplicity~\cite{3GeVLongPaper,3GeVShortPaper,ProtonCumulantRatioOrdering,STAR27and54and200,StarBESICumulantsShortPaper,StarBESICumulants,STARDeuteronCumulants,HadesProtonCumulants,AliceProtonCumulants,AliceAntideuteronCumulants}. We demonstrated the risks associated with using different detectors to measure PID and centrality. The mixed-detector approach enhances the vulnerability of analyses to detector effects, because spontaneous decorrelation between detector responses can cause large high-order cumulants. Pileup corrections to high-order cumulants, which assume that pileup is a superposition of two events, are often invalid in the mixed-detector approach. Attempting to correct for pileup instead of removing it may allow enhanced cumulants to masquerade as a signal for even moderate out-of-time pileup rates.

Spontaneous failure of one detector in a mixed-detector approach, if not carefully removed, will cause fluctuations. The same is true for spontaneous reductions in acceptance, as detector subsystems may reboot during data-taking. This risk is minimized by using the same detector for centrality and PID. 

We do not suggest that the mixed-detector approach should never be used. It is often necessary to use multiple detectors in fluctuations analyses in order to maintain a high proton purity. We emphasize that the mixed-detector approach enhances the vulnerability of analyses to detector effects. Rigorous quality assurance of data is necessary when using multiple detectors to measure high-order cumulants, including aggressive removal of out-of-time pileup. Rare detector failures should be understood and minimized when they do not affect both the multiplicity and PID measurements equally. When detector performance depends on event-by-event algorithms, it is necessary to check, event-by-event, that the detector is performing as expected

Particle identification using multiple detectors is often necessary. When multiple detectors overlap in phase space, one can check the stability of cumulants with respect to how much of each detector is used. When TOF is used for PID for particles above a certain momentum, as in Ref.~\cite{3GeVLongPaper,3GeVShortPaper,ProtonCumulantRatioOrdering}, analyzers may vary the momentum threshold in order to verify the stability of results, as was simulated in the toy models and shown in Figs.~\ref{fig:cumulantsmodel1} and \ref{fig:cumulantsmodel2}. The instability of higher-order cumulants can signal that detector-induced fluctuations are present.

\subsection*{How not to measure a false QCD critical point}

The striking realization from the models shown here is that it is fairly easy to measure a false critical point signature. To be vulnerable to a false signature, an analyzer just needs to use the following two ingredients:
\begin{enumerate}
    \item Measure the multiplicity and proton number with different detectors.
    \item Ensure that the correlated response of the two detectors changes in a small subset of events.
\end{enumerate}
\vspace{2em}

As long as these two actions are taken, an analysis has a high risk of measuring enhanced high-order moments. A large spike in kurtosis at $\sqrt{s_{NN}}$=3.9~GeV, as seen in Fig.~\ref{fig:cumulantsmodel1}, would be interpreted as an exciting confirmation of the predicted behavior of the fourth-order moment near a critical point~\cite{PhysRevLett.107.052301} and of the predicted location of a critical point~\cite{PhysRevD.101.054032, GAO2021136584, PhysRevD.104.054022, refId0, sorensen2024locatingcriticalpointhadron, hippert2023bayesianlocationqcdcritical}. 

The other striking conclusion from these studies is that it is also fairly easy to make a signal that is robust against detector-induced fluctuations. In order to measure a robust signal, an analyzer just needs to follow one guideline:
\begin{enumerate}
    \item Make the multiplicity and proton-number measurements as similar as possible.
\end{enumerate}

Analyzers should maximize the degree to which both measurements are performed by the same detector. They should understand any effects that may cause the correlation between proton number and multiplicity to change. Analyzers should aim for anomalous events to have the same effect on the measured proton number as they do on multiplicity. Otherwise, those rare and spontaneous changes in the correlation between proton number and multiplicity will become the entire measurement.

\section{Acknowledgments}
\label{section:acknowldegments}
We acknowledge useful conversations with Yongcong Xu and Toshihiro Nonaka. This work is supported in part by the U.S. Department of Energy, Office of Science, Office of Nuclear Physics, under contract numbers DE-AC02-05CH11231, by the US National Science Foundation under Grant No. PHY-2209614.


\begin{thebibliography}{26}%
\makeatletter
\providecommand \@ifxundefined [1]{%
 \@ifx{#1\undefined}
}%
\providecommand \@ifnum [1]{%
 \ifnum #1\expandafter \@firstoftwo
 \else \expandafter \@secondoftwo
 \fi
}%
\providecommand \@ifx [1]{%
 \ifx #1\expandafter \@firstoftwo
 \else \expandafter \@secondoftwo
 \fi
}%
\providecommand \natexlab [1]{#1}%
\providecommand \enquote  [1]{``#1''}%
\providecommand \bibnamefont  [1]{#1}%
\providecommand \bibfnamefont [1]{#1}%
\providecommand \citenamefont [1]{#1}%
\providecommand \href@noop [0]{\@secondoftwo}%
\providecommand \href [0]{\begingroup \@sanitize@url \@href}%
\providecommand \@href[1]{\@@startlink{#1}\@@href}%
\providecommand \@@href[1]{\endgroup#1\@@endlink}%
\providecommand \@sanitize@url [0]{\catcode `\\12\catcode `\$12\catcode
  `\&12\catcode `\#12\catcode `\^12\catcode `\_12\catcode `\%12\relax}%
\providecommand \@@startlink[1]{}%
\providecommand \@@endlink[0]{}%
\providecommand \url  [0]{\begingroup\@sanitize@url \@url }%
\providecommand \@url [1]{\endgroup\@href {#1}{\urlprefix }}%
\providecommand \urlprefix  [0]{URL }%
\providecommand \Eprint [0]{\href }%
\providecommand \doibase [0]{https://doi.org/}%
\providecommand \selectlanguage [0]{\@gobble}%
\providecommand \bibinfo  [0]{\@secondoftwo}%
\providecommand \bibfield  [0]{\@secondoftwo}%
\providecommand \translation [1]{[#1]}%
\providecommand \BibitemOpen [0]{}%
\providecommand \bibitemStop [0]{}%
\providecommand \bibitemNoStop [0]{.\EOS\space}%
\providecommand \EOS [0]{\spacefactor3000\relax}%
\providecommand \BibitemShut  [1]{\csname bibitem#1\endcsname}%
\let\auto@bib@innerbib\@empty
\bibitem [{\citenamefont {Stephanov}(2009)}]{PhysRevLett.102.032301}%
  \BibitemOpen
  \bibfield  {author} {\bibinfo {author} {\bibfnamefont {M.~A.}\ \bibnamefont
  {Stephanov}},\ }\href {https://doi.org/10.1103/PhysRevLett.102.032301}
  {\bibfield  {journal} {\bibinfo  {journal} {Phys. Rev. Lett.}\ }\textbf
  {\bibinfo {volume} {102}},\ \bibinfo {pages} {032301} (\bibinfo {year}
  {2009})}\BibitemShut {NoStop}%
\bibitem [{\citenamefont {Stephanov}(2011{\natexlab{a}})}]{Stephanov_2011}%
  \BibitemOpen
  \bibfield  {author} {\bibinfo {author} {\bibfnamefont {M.~A.}\ \bibnamefont
  {Stephanov}},\ }\href {https://doi.org/10.1088/0954-3899/38/12/124147}
  {\bibfield  {journal} {\bibinfo  {journal} {Journal of Physics G: Nuclear and
  Particle Physics}\ }\textbf {\bibinfo {volume} {38}},\ \bibinfo {pages}
  {124147} (\bibinfo {year} {2011}{\natexlab{a}})}\BibitemShut {NoStop}%
\bibitem [{\citenamefont
  {Stephanov}(2011{\natexlab{b}})}]{PhysRevLett.107.052301}%
  \BibitemOpen
  \bibfield  {author} {\bibinfo {author} {\bibfnamefont {M.~A.}\ \bibnamefont
  {Stephanov}},\ }\href {https://doi.org/10.1103/PhysRevLett.107.052301}
  {\bibfield  {journal} {\bibinfo  {journal} {Phys. Rev. Lett.}\ }\textbf
  {\bibinfo {volume} {107}},\ \bibinfo {pages} {052301} (\bibinfo {year}
  {2011}{\natexlab{b}})}\BibitemShut {NoStop}%
\bibitem [{\citenamefont {Abdallah}\ \emph {et~al.}(2023)\citenamefont
  {Abdallah} \emph {et~al.}}]{3GeVLongPaper}%
  \BibitemOpen
  \bibfield  {author} {\bibinfo {author} {\bibfnamefont {M.~S.}\ \bibnamefont
  {Abdallah}} \emph {et~al.} (\bibinfo {collaboration} {STAR Collaboration}),\
  }\href {https://doi.org/10.1103/PhysRevC.107.024908} {\bibfield  {journal}
  {\bibinfo  {journal} {Phys. Rev. C}\ }\textbf {\bibinfo {volume} {107}},\
  \bibinfo {pages} {024908} (\bibinfo {year} {2023})}\BibitemShut {NoStop}%
\bibitem [{\citenamefont {Abdallah}\ \emph {et~al.}(2022)\citenamefont
  {Abdallah} \emph {et~al.}}]{3GeVShortPaper}%
  \BibitemOpen
  \bibfield  {author} {\bibinfo {author} {\bibfnamefont {M.~S.}\ \bibnamefont
  {Abdallah}} \emph {et~al.} (\bibinfo {collaboration} {STAR Collaboration}),\
  }\href {https://doi.org/10.1103/PhysRevLett.128.202303} {\bibfield  {journal}
  {\bibinfo  {journal} {Phys. Rev. Lett.}\ }\textbf {\bibinfo {volume} {128}},\
  \bibinfo {pages} {202303} (\bibinfo {year} {2022})}\BibitemShut {NoStop}%
\bibitem [{\citenamefont {Aboona}\ \emph {et~al.}(2023)\citenamefont {Aboona}
  \emph {et~al.}}]{ProtonCumulantRatioOrdering}%
  \BibitemOpen
  \bibfield  {author} {\bibinfo {author} {\bibfnamefont {B.~E.}\ \bibnamefont
  {Aboona}} \emph {et~al.} (\bibinfo {collaboration} {STAR Collaboration}),\
  }\href {https://doi.org/10.1103/PhysRevLett.130.082301} {\bibfield  {journal}
  {\bibinfo  {journal} {Phys. Rev. Lett.}\ }\textbf {\bibinfo {volume} {130}},\
  \bibinfo {pages} {082301} (\bibinfo {year} {2023})}\BibitemShut {NoStop}%
\bibitem [{\citenamefont {Abdallah}\ \emph
  {et~al.}(2021{\natexlab{a}})\citenamefont {Abdallah} \emph
  {et~al.}}]{STAR27and54and200}%
  \BibitemOpen
  \bibfield  {author} {\bibinfo {author} {\bibfnamefont {M.~S.}\ \bibnamefont
  {Abdallah}} \emph {et~al.} (\bibinfo {collaboration} {STAR Collaboration}),\
  }\href {https://doi.org/10.1103/PhysRevLett.127.262301} {\bibfield  {journal}
  {\bibinfo  {journal} {Phys. Rev. Lett.}\ }\textbf {\bibinfo {volume} {127}},\
  \bibinfo {pages} {262301} (\bibinfo {year} {2021}{\natexlab{a}})}\BibitemShut
  {NoStop}%
\bibitem [{\citenamefont {Adam}\ \emph {et~al.}(2021)\citenamefont {Adam} \emph
  {et~al.}}]{StarBESICumulantsShortPaper}%
  \BibitemOpen
  \bibfield  {author} {\bibinfo {author} {\bibfnamefont {J.}~\bibnamefont
  {Adam}} \emph {et~al.} (\bibinfo {collaboration} {STAR Collaboration}),\
  }\href {https://doi.org/10.1103/PhysRevLett.126.092301} {\bibfield  {journal}
  {\bibinfo  {journal} {Phys. Rev. Lett.}\ }\textbf {\bibinfo {volume} {126}},\
  \bibinfo {pages} {092301} (\bibinfo {year} {2021})}\BibitemShut {NoStop}%
\bibitem [{\citenamefont {Abdallah}\ \emph
  {et~al.}(2021{\natexlab{b}})\citenamefont {Abdallah} \emph
  {et~al.}}]{StarBESICumulants}%
  \BibitemOpen
  \bibfield  {author} {\bibinfo {author} {\bibfnamefont {M.~S.}\ \bibnamefont
  {Abdallah}} \emph {et~al.} (\bibinfo {collaboration} {STAR Collaboration}),\
  }\href {https://doi.org/10.1103/PhysRevC.104.024902} {\bibfield  {journal}
  {\bibinfo  {journal} {Phys. Rev. C}\ }\textbf {\bibinfo {volume} {104}},\
  \bibinfo {pages} {024902} (\bibinfo {year} {2021}{\natexlab{b}})}\BibitemShut
  {NoStop}%
\bibitem [{\citenamefont {Abdulhamid}\ \emph {et~al.}(2024)\citenamefont
  {Abdulhamid} \emph {et~al.}}]{STARDeuteronCumulants}%
  \BibitemOpen
  \bibfield  {author} {\bibinfo {author} {\bibfnamefont {M.}~\bibnamefont
  {Abdulhamid}} \emph {et~al.} (\bibinfo {collaboration} {STAR
  Collaboration}),\ }\href
  {https://doi.org/https://doi.org/10.1016/j.physletb.2024.138560} {\bibfield
  {journal} {\bibinfo  {journal} {Physics Letters B}\ }\textbf {\bibinfo
  {volume} {855}},\ \bibinfo {pages} {138560} (\bibinfo {year}
  {2024})}\BibitemShut {NoStop}%
\bibitem [{\citenamefont {Adamczewski-Musch}\ \emph {et~al.}(2020)\citenamefont
  {Adamczewski-Musch} \emph {et~al.}}]{HadesProtonCumulants}%
  \BibitemOpen
  \bibfield  {author} {\bibinfo {author} {\bibfnamefont {J.}~\bibnamefont
  {Adamczewski-Musch}} \emph {et~al.} (\bibinfo {collaboration} {HADES
  Collaboration}),\ }\href {https://doi.org/10.1103/PhysRevC.102.024914}
  {\bibfield  {journal} {\bibinfo  {journal} {Phys. Rev. C}\ }\textbf {\bibinfo
  {volume} {102}},\ \bibinfo {pages} {024914} (\bibinfo {year}
  {2020})}\BibitemShut {NoStop}%
\bibitem [{\citenamefont {Acharya}\ \emph
  {et~al.}(2023{\natexlab{a}})\citenamefont {Acharya} \emph
  {et~al.}}]{AliceProtonCumulants}%
  \BibitemOpen
  \bibfield  {author} {\bibinfo {author} {\bibfnamefont {S.}~\bibnamefont
  {Acharya}} \emph {et~al.} (\bibinfo {collaboration} {ALICE Collaboration}),\
  }\href {https://doi.org/https://doi.org/10.1016/j.physletb.2022.137545}
  {\bibfield  {journal} {\bibinfo  {journal} {Physics Letters B}\ }\textbf
  {\bibinfo {volume} {844}},\ \bibinfo {pages} {137545} (\bibinfo {year}
  {2023}{\natexlab{a}})}\BibitemShut {NoStop}%
\bibitem [{\citenamefont {Acharya}\ \emph
  {et~al.}(2023{\natexlab{b}})\citenamefont {Acharya} \emph
  {et~al.}}]{AliceAntideuteronCumulants}%
  \BibitemOpen
  \bibfield  {author} {\bibinfo {author} {\bibfnamefont {S.}~\bibnamefont
  {Acharya}} \emph {et~al.} (\bibinfo {collaboration} {ALICE Collaboration}),\
  }\href {https://doi.org/10.1103/PhysRevLett.131.041901} {\bibfield  {journal}
  {\bibinfo  {journal} {Phys. Rev. Lett.}\ }\textbf {\bibinfo {volume} {131}},\
  \bibinfo {pages} {041901} (\bibinfo {year} {2023}{\natexlab{b}})}\BibitemShut
  {NoStop}%
\bibitem [{\citenamefont {Fu}\ \emph {et~al.}(2020)\citenamefont {Fu},
  \citenamefont {Pawlowski},\ and\ \citenamefont
  {Rennecke}}]{PhysRevD.101.054032}%
  \BibitemOpen
  \bibfield  {author} {\bibinfo {author} {\bibfnamefont {W.-j.}\ \bibnamefont
  {Fu}}, \bibinfo {author} {\bibfnamefont {J.~M.}\ \bibnamefont {Pawlowski}},\
  and\ \bibinfo {author} {\bibfnamefont {F.}~\bibnamefont {Rennecke}},\ }\href
  {https://doi.org/10.1103/PhysRevD.101.054032} {\bibfield  {journal} {\bibinfo
   {journal} {Phys. Rev. D}\ }\textbf {\bibinfo {volume} {101}},\ \bibinfo
  {pages} {054032} (\bibinfo {year} {2020})}\BibitemShut {NoStop}%
\bibitem [{\citenamefont {Gao}\ and\ \citenamefont
  {Pawlowski}(2021)}]{GAO2021136584}%
  \BibitemOpen
  \bibfield  {author} {\bibinfo {author} {\bibfnamefont {F.}~\bibnamefont
  {Gao}}\ and\ \bibinfo {author} {\bibfnamefont {J.~M.}\ \bibnamefont
  {Pawlowski}},\ }\href
  {https://doi.org/https://doi.org/10.1016/j.physletb.2021.136584} {\bibfield
  {journal} {\bibinfo  {journal} {Physics Letters B}\ }\textbf {\bibinfo
  {volume} {820}},\ \bibinfo {pages} {136584} (\bibinfo {year}
  {2021})}\BibitemShut {NoStop}%
\bibitem [{\citenamefont {Gunkel}\ and\ \citenamefont
  {Fischer}(2021)}]{PhysRevD.104.054022}%
  \BibitemOpen
  \bibfield  {author} {\bibinfo {author} {\bibfnamefont {P.~J.}\ \bibnamefont
  {Gunkel}}\ and\ \bibinfo {author} {\bibfnamefont {C.~S.}\ \bibnamefont
  {Fischer}},\ }\href {https://doi.org/10.1103/PhysRevD.104.054022} {\bibfield
  {journal} {\bibinfo  {journal} {Phys. Rev. D}\ }\textbf {\bibinfo {volume}
  {104}},\ \bibinfo {pages} {054022} (\bibinfo {year} {2021})}\BibitemShut
  {NoStop}%
\bibitem [{\citenamefont {{J. Goswami}}\ \emph {et~al.}(2024)\citenamefont {{J.
  Goswami}}, \citenamefont {{D.A. Clarke}}, \citenamefont {{P. Dimopoulos}},
  \citenamefont {{F. Di Renzo}}, \citenamefont {{C. Schmidt}}, \citenamefont
  {{S. Singh}},\ and\ \citenamefont {{K. Zambello}}}]{refId0}%
  \BibitemOpen
  \bibfield  {author} {\bibinfo {author} {\bibnamefont {{J. Goswami}}},
  \bibinfo {author} {\bibnamefont {{D.A. Clarke}}}, \bibinfo {author}
  {\bibnamefont {{P. Dimopoulos}}}, \bibinfo {author} {\bibnamefont {{F. Di
  Renzo}}}, \bibinfo {author} {\bibnamefont {{C. Schmidt}}}, \bibinfo {author}
  {\bibnamefont {{S. Singh}}},\ and\ \bibinfo {author} {\bibnamefont {{K.
  Zambello}}},\ }\href {https://doi.org/10.1051/epjconf/202429606007}
  {\bibfield  {journal} {\bibinfo  {journal} {EPJ Web Conf.}\ }\textbf
  {\bibinfo {volume} {296}},\ \bibinfo {pages} {06007} (\bibinfo {year}
  {2024})}\BibitemShut {NoStop}%
\bibitem [{\citenamefont {Sorensen}\ and\ \citenamefont
  {Sorensen}(2024)}]{sorensen2024locatingcriticalpointhadron}%
  \BibitemOpen
  \bibfield  {author} {\bibinfo {author} {\bibfnamefont {A.}~\bibnamefont
  {Sorensen}}\ and\ \bibinfo {author} {\bibfnamefont {P.}~\bibnamefont
  {Sorensen}},\ }\href {https://arxiv.org/abs/2405.10278} {} (\bibinfo {year}
  {2024}),\ \Eprint {https://arxiv.org/abs/2405.10278} {arXiv:2405.10278
  [nucl-th]} \BibitemShut {NoStop}%
\bibitem [{\citenamefont {Hippert}\ \emph {et~al.}(2023)\citenamefont {Hippert}
  \emph {et~al.}}]{hippert2023bayesianlocationqcdcritical}%
  \BibitemOpen
  \bibfield  {author} {\bibinfo {author} {\bibfnamefont {M.}~\bibnamefont
  {Hippert}} \emph {et~al.},\ }\href {https://arxiv.org/abs/2309.00579} {}
  (\bibinfo {year} {2023}),\ \Eprint {https://arxiv.org/abs/2309.00579}
  {arXiv:2309.00579 [nucl-th]} \BibitemShut {NoStop}%
\bibitem [{201(2019)}]{2019TheSB}%
  \BibitemOpen
  \href {https://api.semanticscholar.org/CorpusID:220267203} {\bibinfo {title}
  {The star beam use request for run-20 and run-21}} (\bibinfo {year}
  {2019})\BibitemShut {NoStop}%
\bibitem [{\citenamefont {Braun-Munzinger}\ \emph {et~al.}(2017)\citenamefont
  {Braun-Munzinger}, \citenamefont {Rustamov},\ and\ \citenamefont
  {Stachel}}]{BRAUNMUNZINGER2017114}%
  \BibitemOpen
  \bibfield  {author} {\bibinfo {author} {\bibfnamefont {P.}~\bibnamefont
  {Braun-Munzinger}}, \bibinfo {author} {\bibfnamefont {A.}~\bibnamefont
  {Rustamov}},\ and\ \bibinfo {author} {\bibfnamefont {J.}~\bibnamefont
  {Stachel}},\ }\href
  {https://doi.org/https://doi.org/10.1016/j.nuclphysa.2017.01.011} {\bibfield
  {journal} {\bibinfo  {journal} {Nuclear Physics A}\ }\textbf {\bibinfo
  {volume} {960}},\ \bibinfo {pages} {114} (\bibinfo {year}
  {2017})}\BibitemShut {NoStop}%
\bibitem [{\citenamefont {Skokov}\ \emph {et~al.}(2013)\citenamefont {Skokov},
  \citenamefont {Friman},\ and\ \citenamefont {Redlich}}]{PhysRevC.88.034911}%
  \BibitemOpen
  \bibfield  {author} {\bibinfo {author} {\bibfnamefont {V.}~\bibnamefont
  {Skokov}}, \bibinfo {author} {\bibfnamefont {B.}~\bibnamefont {Friman}},\
  and\ \bibinfo {author} {\bibfnamefont {K.}~\bibnamefont {Redlich}},\ }\href
  {https://doi.org/10.1103/PhysRevC.88.034911} {\bibfield  {journal} {\bibinfo
  {journal} {Phys. Rev. C}\ }\textbf {\bibinfo {volume} {88}},\ \bibinfo
  {pages} {034911} (\bibinfo {year} {2013})}\BibitemShut {NoStop}%
\bibitem [{\citenamefont {Nonaka}\ \emph {et~al.}(2017)\citenamefont {Nonaka},
  \citenamefont {Kitazawa},\ and\ \citenamefont {Esumi}}]{PhysRevC.95.064912}%
  \BibitemOpen
  \bibfield  {author} {\bibinfo {author} {\bibfnamefont {T.}~\bibnamefont
  {Nonaka}}, \bibinfo {author} {\bibfnamefont {M.}~\bibnamefont {Kitazawa}},\
  and\ \bibinfo {author} {\bibfnamefont {S.}~\bibnamefont {Esumi}},\ }\href
  {https://doi.org/10.1103/PhysRevC.95.064912} {\bibfield  {journal} {\bibinfo
  {journal} {Phys. Rev. C}\ }\textbf {\bibinfo {volume} {95}},\ \bibinfo
  {pages} {064912} (\bibinfo {year} {2017})}\BibitemShut {NoStop}%
\bibitem [{\citenamefont {Ackermann}\ \emph {et~al.}(2003)\citenamefont
  {Ackermann} \emph {et~al.}}]{STAR:2002eio}%
  \BibitemOpen
  \bibfield  {author} {\bibinfo {author} {\bibfnamefont {K.~H.}\ \bibnamefont
  {Ackermann}} \emph {et~al.} (\bibinfo {collaboration} {STAR Collaboration}),\
  }\href {https://doi.org/10.1016/S0168-9002(02)01960-5} {\bibfield  {journal}
  {\bibinfo  {journal} {Nucl. Instrum. Meth. A}\ }\textbf {\bibinfo {volume}
  {499}},\ \bibinfo {pages} {624} (\bibinfo {year} {2003})}\BibitemShut
  {NoStop}%
\bibitem [{\citenamefont {Nonaka}\ \emph {et~al.}(2020)\citenamefont {Nonaka},
  \citenamefont {Kitazawa},\ and\ \citenamefont {Esumi}}]{NONAKA2020164632}%
  \BibitemOpen
  \bibfield  {author} {\bibinfo {author} {\bibfnamefont {T.}~\bibnamefont
  {Nonaka}}, \bibinfo {author} {\bibfnamefont {M.}~\bibnamefont {Kitazawa}},\
  and\ \bibinfo {author} {\bibfnamefont {S.}~\bibnamefont {Esumi}},\ }\href
  {https://doi.org/https://doi.org/10.1016/j.nima.2020.164632} {\bibfield
  {journal} {\bibinfo  {journal} {Nucl. Instrum. Meth. A}\ }\textbf {\bibinfo
  {volume} {984}},\ \bibinfo {pages} {164632} (\bibinfo {year}
  {2020})}\BibitemShut {NoStop}%
\bibitem [{\citenamefont {Zhang}\ \emph {et~al.}(2022)\citenamefont {Zhang},
  \citenamefont {Huang}, \citenamefont {Nonaka},\ and\ \citenamefont
  {Luo}}]{ZHANG2022166246}%
  \BibitemOpen
  \bibfield  {author} {\bibinfo {author} {\bibfnamefont {Y.}~\bibnamefont
  {Zhang}}, \bibinfo {author} {\bibfnamefont {Y.}~\bibnamefont {Huang}},
  \bibinfo {author} {\bibfnamefont {T.}~\bibnamefont {Nonaka}},\ and\ \bibinfo
  {author} {\bibfnamefont {X.}~\bibnamefont {Luo}},\ }\href
  {https://doi.org/https://doi.org/10.1016/j.nima.2021.166246} {\bibfield
  {journal} {\bibinfo  {journal} {Nucl. Instrum. Meth. A}\ }\textbf {\bibinfo
  {volume} {1026}},\ \bibinfo {pages} {166246} (\bibinfo {year}
  {2022})}\BibitemShut {NoStop}%
\end{thebibliography}
\end{document}